\documentclass[aps,twocolumn,prd,showpacs,preprintnumbers,nofootinbib,showpacs,preprintnumbers,amsmath,amssymb,aps,prd,superscriptaddress]{revtex4-1}

\usepackage{graphicx}
\usepackage{dcolumn}
\usepackage{bm}
\usepackage{hyperref}
\usepackage{color}
\usepackage{float}
\usepackage{ulem}

\usepackage[export]{adjustbox}
\usepackage[dvipsnames]{xcolor}
\usepackage{ifthen}
\hypersetup{
     colorlinks = true,
     linkcolor = blue,
     anchorcolor = blue,
     citecolor = green,
     filecolor = blue,
     urlcolor = magenta
     }

\def\blue{\textcolor{blue}}

\begin{document}
\title{Chirp mass based glitch identification in long duration gravitational wave transients}

\author{Nirban Bose}
\email{nirban@iitb.ac.in}
\author{Archana Pai}
\email{archana@phy.iitb.ac.in}
\author{Koustav Chandra}
\email{koustav.chandra@iitb.ac.in}
\author{V. Gayathri}
\email{gayathrivignesh@ufl.edu, Currently a post-doctoral fellow at the University of Florida}
\affiliation{Department of Physics, Indian Institute of Technology Bombay, Mumbai, Maharashtra 400076, India}

\begin{abstract}
The data from the ground-based interferometric gravitational wave detectors(GW) is often masqueraded by highly localised short-duration glitches which pose a serious challenge for any transient GW search. In this work, we propose a glitch identification algorithm for the scenario of presence of a glitch along with a long duration GW transient. We device the chirp mass consistency parameter across different Q planes in the Q-transform. We test this algorithm over a variety of short duration glitches from the observing runs along with a long duration binary inspiral signal. The demonstration of  this algorithm for a generic low mass and low eccentricity binary systems shows the  direct application of it in the long duration unmodeled searches. 
\end{abstract}

\maketitle

\section{Introduction}
The first detection of Gravitational Waves(GWs) from merging black holes have opened up the opportunity to study a new family of astrophysical sources~\cite{GW150914-DETECTION}. The LIGO Scientific Collaboration and the Virgo Collaboration(LVC) have detected multiple sources~\cite{GWTC1, GW170817-DETECTION} in the first (O1) and second observing (O2) run by analyzing the data observed by the Advanced LIGO~\cite{ TheLIGOScientific:2014jea} and Virgo~\cite{TheVirgo:2014hva} detectors. KAGRA \cite{Akutsu:2020his} is rapidly reaching sensitivities comparable to LIGO and Virgo and will be soon participating in the upcoming observational runs. The recently concluded third observing (O3) run had approximately $\sim 50$ candidates are reported ~\cite{GraceDB-03-Public} and there has been three confirmed detection ~\cite{Abbott:2020uma, LIGOScientific:2020stg, Abbott_2020}. Several independent analyses have also reported additional candidates based on their study on publicly available data ~\cite{Nitz_2019, Nitz_2020, PhysRevD.100.023011, PhysRevD.101.083030, Antelis:2018smo}. 

With an improvement in the sensitivity of the detectors, the number of detections is expected to increase. However, this improvement also makes the instrument more susceptible to short-duration noise transients, a.k.a. ``glitches," which in turn increases the probability of simultaneous observing a glitch and an astrophysical signal. This has two direct consequences~\cite{Usman:2015kfa}; {\it first}  it increases the false alarms and hence bring down the astrophysical significance of the prospective GW candidate event and {\it second} if we wrongly veto out the segment containing the glitch, it can result in non-detection of the underlying astrophysical GW transient. The latter was the case when the GW from the first binary neutron star (BNS) system (GW170817) was initially vetoed out in LIGO-Livingston because of a glitch that occured just a few seconds prior to the merger ~\cite{GW170817-DETECTION}.

Glitches are a result of  instrumental or environmental artefacts, and the advanced gravitational wave detectors monitor them with  $\sim 10^5$ auxiliary channels~\cite{GW150914-DETECTION}. Despite that, the origin and rate of these glitches are still not well understood~\cite{Cabero_2019,Zevin:2016qwy}. At present there are several methods to identify glitches and most of them rely on the excess power criteria to identify them~\cite{Zhu:2017ujz, Pankow:2018qpo}. 
Here, we consider morphological difference between a glitch when present along with the long duration transient signal to identify the glitch position which can be used to nullify its effect.

Primarily the long duration transients include low mass compact binary signals. While templated searches exist for such sources for circular binaries, no model based searches exist for moderate to high eccentricities. Recent literature suggest that binary systems with appreciable eccentricity are possible in a dense stellar environment like a globular cluster or galactic nuclei~\cite{Rodriguez_2018}. There have been efforts to look for eccentric binary black hole (eBBH)~\cite{Salemi:2019owp} systems in the observed O1-O2 data in a model independent way. Although no eccentric compact binary has been detected so far, rate upper limits are placed on these sources ~\cite{Salemi:2019owp}. In unmodeled searches, presence of glitches can hamper the correct clustering of the excess-power pixels and consequently effect the detection of real astrophysical signals. Hence it is important to find ways to identify glitches and veto them. 
 
In this work, we propose a method that can be used to identify the location of a short duration glitch, for example, a blip glitch, present in the time evolution of a low mass binary system with low to moderate eccentricities.
The proposed method uses excess power criterion, along with consistency of the chirp mass parameter across different Q planes to localize a glitch when superimposed in the chirp track. This along with the excess power method provides a robust criteria for glitch identification in long duration transients for a variety of short duration glitches. For a long duration eBBH unmodeled search, this method can bring in additional merits in glitch identification and subsequent removal of the corresponding pixels to construct the statistic.

This paper is organized as follows : 
in Sec.~\ref{Q-transform in GW search}, we discuss the Q transform response in the GW search. We compute the Q transform response to a chirping GW signal and a short duration glitch to understand their analytical behavior. In Sec.~\ref{Methodology}, we discuss the salient features of the algorithm used for glitch bin identification. We then validate our method by simulating long duration gravitational wave compact binary signals in the detector noise in Sec.~\ref{Simulations}. Finally, we apply our algorithm to the GW170817 data as well as low mass, low eccentric systems in Sec. \ref{Application} and conclude our work in Sec.~\ref{Conclusions}. 

\section{Q-transform in GW search} \label{Q-transform in GW search}
Consider the calibrated detector strain data $x(t)$. The over-whitened data is then given as $\tilde{\tilde{x}}(f) = \tilde{x}(f)/S_n(f)$ where
$\tilde{x}(f)$ is the frequency domain representation of $x(t)$ and $S_n(f)$ is the two sided noise power spectral density (PSD). 

The Q-transform is defined as a windowed Fourier transform that  projects the over-whitened time-series (or frequency series) onto a basis formed by minimum uncertainty waveform~\cite{shouravthesis}. Thus the Q-transform, $X(t,f_0,Q)$ is obtained by acting upon the over-whitened frequency series with an integral operator whose kernel is $\mathfrak{\tilde{w}^*}(f,f_0,Q)e^{i2\pi f t}$, as \cite{Chatterji_2004}:
\begin{equation}\label{eq:1}
\begin{split}
    X(t,f_0,Q) 
    &= \int_{-\infty}^{\infty} df ~ \tilde{\tilde{x}}^*(f+f_0)\mathfrak{\tilde{w}}(f,f_0,Q)e^{i2\pi f t}
    \end{split}
\end{equation}
Here $f_0$ is the central frequency and $Q$ is the Q-tile used to calculate the Q-transform. The width of the window function in time-domain is $\tau$. The window function in frequency domain $\mathfrak{\tilde{w}}(f,f_0,Q)$ is a normalised bi-square window with finite frequency domain support centered at zero and it is given as \cite{shouravthesis} :
\begin{equation}\label{eq:2}
    \mathfrak{\tilde{w}}(f,f_0,Q) = \begin{cases}
\sqrt{\frac{315Q}{128\sqrt{11}f_0}}\Big[1 - \Big(\frac{fQ}{f_0\sqrt{11}}\Big)^2 \Big]^2, \hfill \|f\| \leq \frac{f_0\sqrt{11}}{Q} \\
\\
0, \hfill otherwise
\end{cases}
\end{equation}

A discrete Q-transform gives a multi-resolution time-frequency map where the time and frequency resolution depends on the $Q$ value (for details see, e.g.,~\cite{Chatterji_2004}).

\subsection{Q-transform of GW chirp}
\label{Q-transform of GW chirp}
To Newtonian order, the GW signal, in the frequency domain, emitted by a quasi-periodic slow moving binary is given as \cite{Maggiore} :
\begin{equation}
    \label{eq:3}
    \begin{split}
        \tilde{h}(f) &=  \sqrt{\frac{5\pi}{24}}\frac{G^2\mathcal{M}^2}{c^5D_L}(\pi G \mathcal{M} f/c^3)^{-7/6}e^{-i\Psi(f)} \\
        &\equiv h_0 \mathcal{M}^{5/6} f^{-7/6} e^{-i\Psi(f)}
    \end{split}
\end{equation}where $\mathcal{M} = (m_1m_2)^{3/5}/(m_1 + m_2)^{1/5}$ is the chirp mass of the system whose component masses are $m_1$ and $m_2$. Here $D_L$ is the luminosity distance. The effective phase of such a Newtonian chirp, $\Psi(f)$, is given as \cite{Maggiore} :
\begin{equation}\label{eq:4}
    \Psi(f) = 2\pi ft_c - 2 \phi_c - \frac{\pi}{4} + \frac{3}{128} (\pi G \mathcal{M} f/c^3)^{-5/3}
\end{equation}
with $t_c$ and $\phi_c$  being the coalescence time and the phase at coalescence.

To calculate the Q-transform of the Newtonian chirp, we approximate the bi-square window with a Gaussian window of width $2f_0/Q$. Clearly, for small values of $Q$, the frequency/time domain width is larger/smaller. This is not optimal as chirp spends a longer time at lower frequencies. Hence, we expect to obtain the maximum signal energy at large values of Q where the width in frequency domain will be small.

\begin{figure}[h!]
   \centering
    \includegraphics[width=\columnwidth]{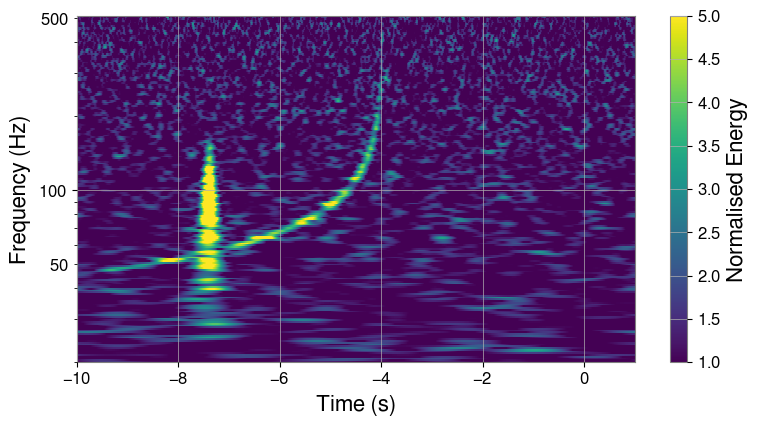}
    \caption{Q transform of an injected signal of 1.4$M_{\odot}$-1.4$M_{\odot}$ in presence of a blip.}
    \label{Q transform of signal}
\end{figure}

However, with further increase in Q, the total energy drops as the time window is much larger than required to capture the frequency content. In Appendix~\ref{A1}, we analytically compute the Q-transform of the Newtonian chirp which captures this feature as given below:
\begin{equation}
 \|X_{chirp}\| \propto \frac{h_0 \mathcal{M}_c^{5/6}}{\sqrt{Q}} \frac{f_0^{-2/3}}{S_n(f_0)}  \exp \left(\frac{- K^2 f_0^2} {2 Q^2} \right)
    \label{Q transform signal}
\end{equation}
Here, $K=0$ corresponds to the primary chirp path in the Q-plane. The brightest pixel at around 100Hz in Fig.~\ref{Q transform of signal} correspond to the most sensitive frequency in the noise PSD of the interferometer.

\subsection{Q-transform of short duration glitch}
\label{Q-transform of noisy blip}

\begin{figure}[h!]
    \centering
    \includegraphics[width=\columnwidth]{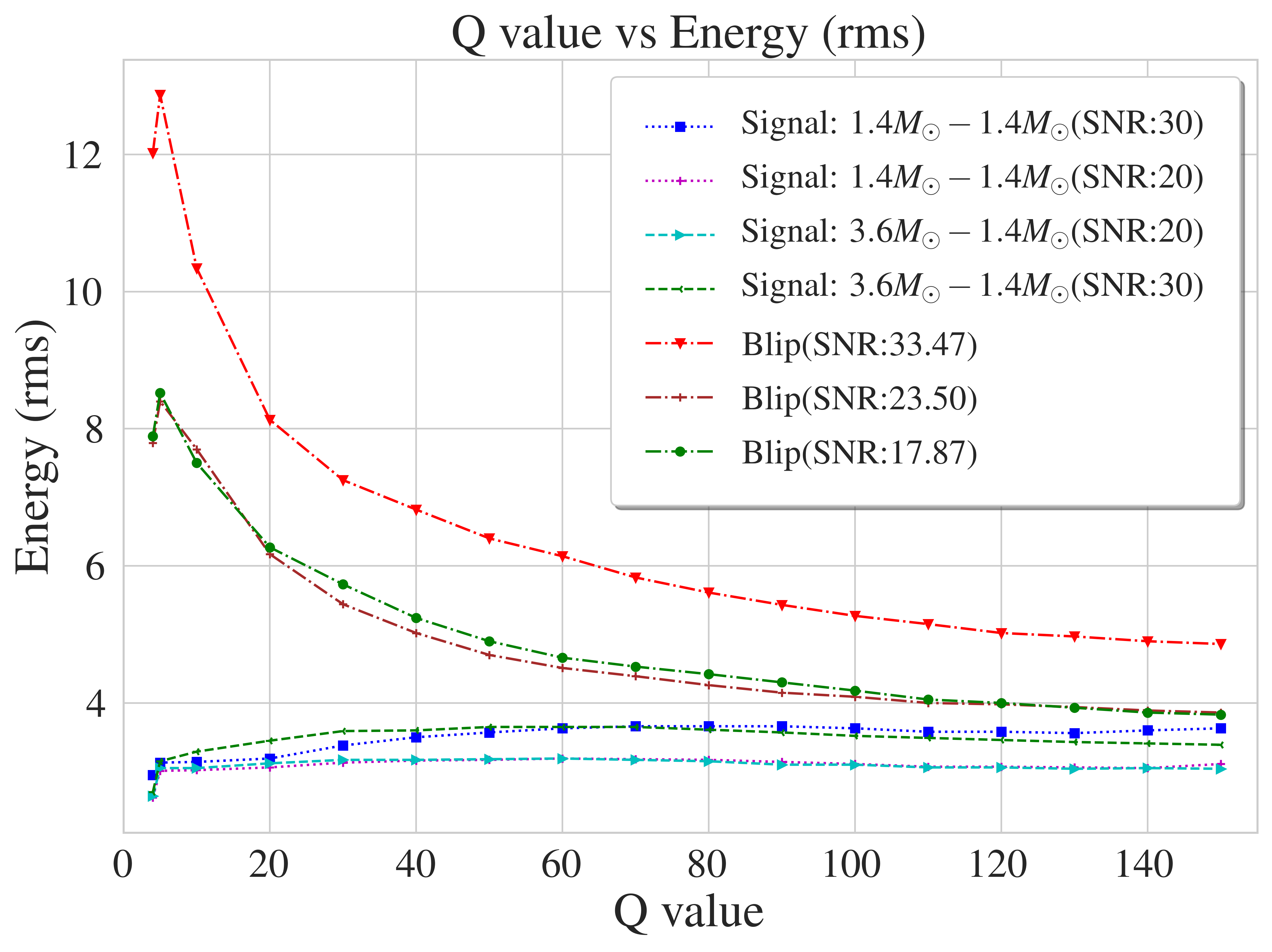}
    \caption{Energy variation of signals and glitches with respect to different Q values}
    \label{Energy variation with respect to different Q values}
\end{figure}

To understand the analytical behaviour of glitches in a TF map, we evaluate the Q-transform. The short noisy transients are well-localised in time but can have a larger frequency bandwidth~\cite{Cabero:2019orq, gravityspy}. They can be reasonably modeled by a sine-Gaussian function with a central frequency $f_{sg}$ and decay time $t_{sg}$. As shown in Appendix~\ref{appendixB}, the Q-transform of such a sine-Gaussian blip glitch is given as,
 \begin{equation}
 \begin{split}
\| X \| \propto \frac{\tau_{sg}}{S_n(f_0)} \sqrt{\frac{f_0}{Q}}  \exp \Big[- \pi^2 (\tau_{sg}^2 (f_0-f_{sg})^2 - \\
\pi^2 (t - t_{sg})^2 \delta^2\Big]
\end{split}
\label{eq:Blip}
\end{equation}
This shows that for a fixed $Q$, the Q-transform peaks when $(f_0,t)$ coincides with $(f_{sg},t_{sg})$ as shown in Fig.~\ref{Q-transform of noisy blip}. This value drops with an increase in Q as shown in Fig.\ref{Energy variation with respect to different Q values}

\begin{widetext}
\begin{figure*}[t!]
\hskip2.0cm
\begin{tabular}{ccc}
\centering
\includegraphics[width=60mm]{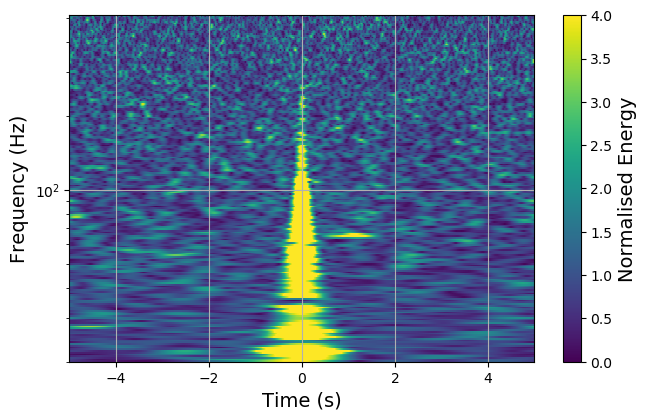}& 
\includegraphics[width=60mm]{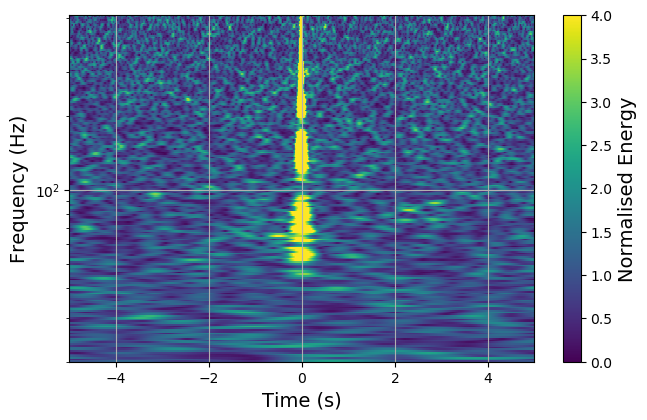} &
\includegraphics[width=60mm]{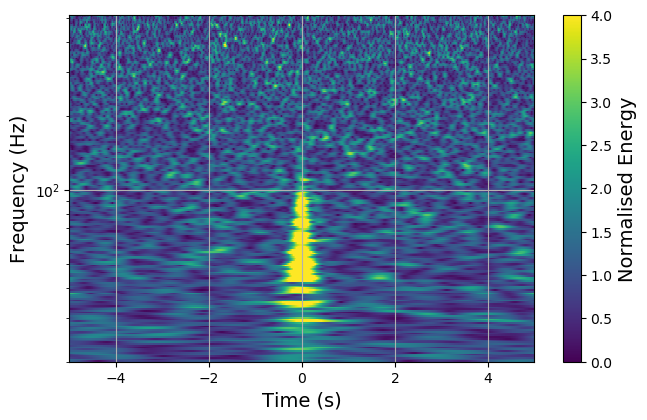}\\
(a) Fat blip & (b) Blip with horn & (c) Tomte blip 
\end{tabular}
\caption{Q transforms of some different types of blips from O1-O2 data used for our study.}
\label{diff types of blips}
\end{figure*}
\end{widetext}

\subsection{Example}
In this subsection, we study the Q-transform of signal and short duration glitch using PyCBC~\cite{Usman:2015kfa}. In Fig.~\ref{Energy variation with respect to different Q values}, we plot the energy as a function of the Q-value for both signal and the blip. Here the energy corresponds to the total RMS energy of the signal (glitch) normalized by the number of pixels having excess energy and is proportional to $ \|X_{chirp}\|^2$ ($\|X_{sg}\|^2$).

 We choose three different short noise transients having SNRs - 33.47, 23.50, 17.87 from O2 data and compute the Q-transform.\footnote{Here, signal SNR refers to optimal matched filter SNR and blip SNR refers to the SNR obtained from the matched filter template.} We observe that for a glitch, the RMS energy decreases with an increase in  Q and they are well captured by $Q \sim 5$. This is because they are well localised in time. For the signal case, we inject long duration signals of SNRs 20 and 30 in O2 noise and compute the Q transform.  The RMS energy of the signals increases with Q and flattens saturate quickly beyond $ Q \sim 30$. After that, with an increase in Q, the energy gradually starts to decrease at $Q > 100$. Both the trends are consistent with the analytical calculations presented in the Appendix.

\section{Algorithm}
\label{Methodology}
The Q transform shows a distinct response to the gravitational wave chirps and noisy glitches. In the case of long duration GW chirp, we can extract the chirp mass from the frequency evolution track in the Q-transform. The chirp mass obtained is not expected to change for small change in Q-tile values.  The noisy glitch on the otherhand does not hold this property. This motivates us to use Q-transform as a tool to localize the glitch location in the underlying GW transient with a super-posed noisy glitch. Below, we lay down the proposed algorithm.

\subsection{Chirp mass variation with respect to Q-plane}\label{Chirp mass estimation}
In the Newtonian order, the frequency evolution of a gravitational waves from a binary inspiral is related to the chirp mass $\mathcal{M}$ as, 
\begin{equation}
    \dot{f}=\frac{96 \pi^{8/3}}{5} {\left(\frac{G\mathcal{M}}{c^3} \right)}^{5/3} f^{11/3}
    \label{f dot}
\end{equation}
Eqn~\eqref{f dot} when integrated over time yields, 
\begin{equation}
    \frac{96 \pi^{8/3}}{5} {\left(\frac{G\mathcal{M}}{c^3}\right)}^{5/3} t +  \frac{3}{8} f^{-8/3} + C  = 0
    \label{f_8/3 vs t}
\end{equation}
where $C$ is the integration constant. The chirp mass $\mathcal{M}$ can be estimated from Eqn~\eqref{f_8/3 vs t} after extracting the time and frequency content of the event~\cite{Abbott:2018wiz}. To see how well our data fits the curve, we compute the reduced chi square statistic $\chi^2_r =\frac{1}{\nu} \sum_{i} \frac{{(y_{obs,i} - y_{fit,i})}^2}{\sigma_i^2}$ , where $y_{obs,i}, y_{fit,i}$ are the observed and fitted data points respectively and  $\nu$ denotes the degrees of freedom. The systematic error $\sigma_i$ is given by
\begin{equation} 
\sigma_i^2 = {({f_i^{-11/3}}\Delta f_i)}^2 + {(b \Delta t_i)}^2 
\end{equation}
where b is the slope of Eqn~\eqref{f_8/3 vs t}, and $\Delta f_i, \Delta t_i $ being the frequency and time resolution of the $i^{th}$  pixel.

 For the gravitational wave from low mass coalescing compact binary merger (and hence with a significant inspiral part), such a chirp mass estimation will be closer to its actual value~\cite{Abbott:2018wiz} unlike the blip. The chirp mass parameter obtained from time-frequency evolution of the blip deviates significantly from the chirp mass of the template it clicked. 
 
We show in the Sec.~\ref{Q-transform in GW search}, the projected signal energy in the Q plane depends on the nature of the signal as given in Eqs.(\ref{Q transform signal}-\ref{eq:Blip}). It is clear that the Q plane values should be wisely chosen such that the energy of the blip, as well as a binary signal, is sufficient for obtaining the chirp track and subsequently estimating the chirp mass. We choose a Q-value in the range 80-120 as the variation in energy of both glitch and signal is minimum in this range (See Fig.~\ref{Energy variation with respect to different Q values}).

 For typical long-duration coalescing compact binary merger signals, the estimated chirp mass across different Q planes ranging from 80-120 is consistent within 95$\%$ value of the actual chirp mass. At the same time, for the blips, it varies between 20$\%$ to about 90$\%$ value of the chirp mass obtained by the best Q plane in (80-120) range.  We use this consistency of chirp mass across the different neighboring Q planes to find the position of the blip bin when coexisting with the long duration chirpy transient.

\subsection{Glitch bin identification}
\label{Blip bin identification}
We begin by evaluating the Q-transform of a given data segment in the aforementioned Q-range and choose the one which maximises the median projected energy; $Q_{best}$. We then divide the Q-transform data into segments of size 0.5s.

We then select the bright pixels which are above a threshold pixel energy value. A stringent cut would miss parts of the signal or glitch, while too much conservative value would introduce noisy pixels. Further, we use the distribution of frequency of the selected pixels in each time bin to eliminate noisy pixels in each bin by removing pixels in the tail. 

After the primary pixel selection process, we make use of the properties of signal to identify glitch location if present: {\it First}, as seen in~\ref{Chirp mass estimation}, the chirp mass estimation with the signal is consistent amongst a plane neighboring to $Q_{best}$. In contrast, for the glitch the chirp mass parameter varies to a considerable extent for a given bin. This motivates us to choose the deviation of the difference in chirp mass estimation from the neighboring Q planes as a consistency test parameter to identify the glitch as against the signal bins.

\begin{equation}
D \equiv \Delta {\cal M}_+ -  \Delta {\cal M}_- \,.
\end{equation}
Here $\Delta {\cal M}_\pm = |{\cal M}_{Q_{best}} - {\cal M}_{Q_\pm} |$ 
is the absolute difference between chirp mass estimation between $Q_{best}$ and a plane neighboring to $Q_{best}$. The {\it second} property we use is the localization of a short duration glitch which will result in a large number of pixels in the glitch bin except for the merger time bin of a signal with high SNR. 

\section{Simulations}
\label{Simulations}
In this section, we validate the proposed method by simulations. 
\subsection{Validation}
\label{Validation}

\begin{figure}[h!]
    \includegraphics[width=\columnwidth]{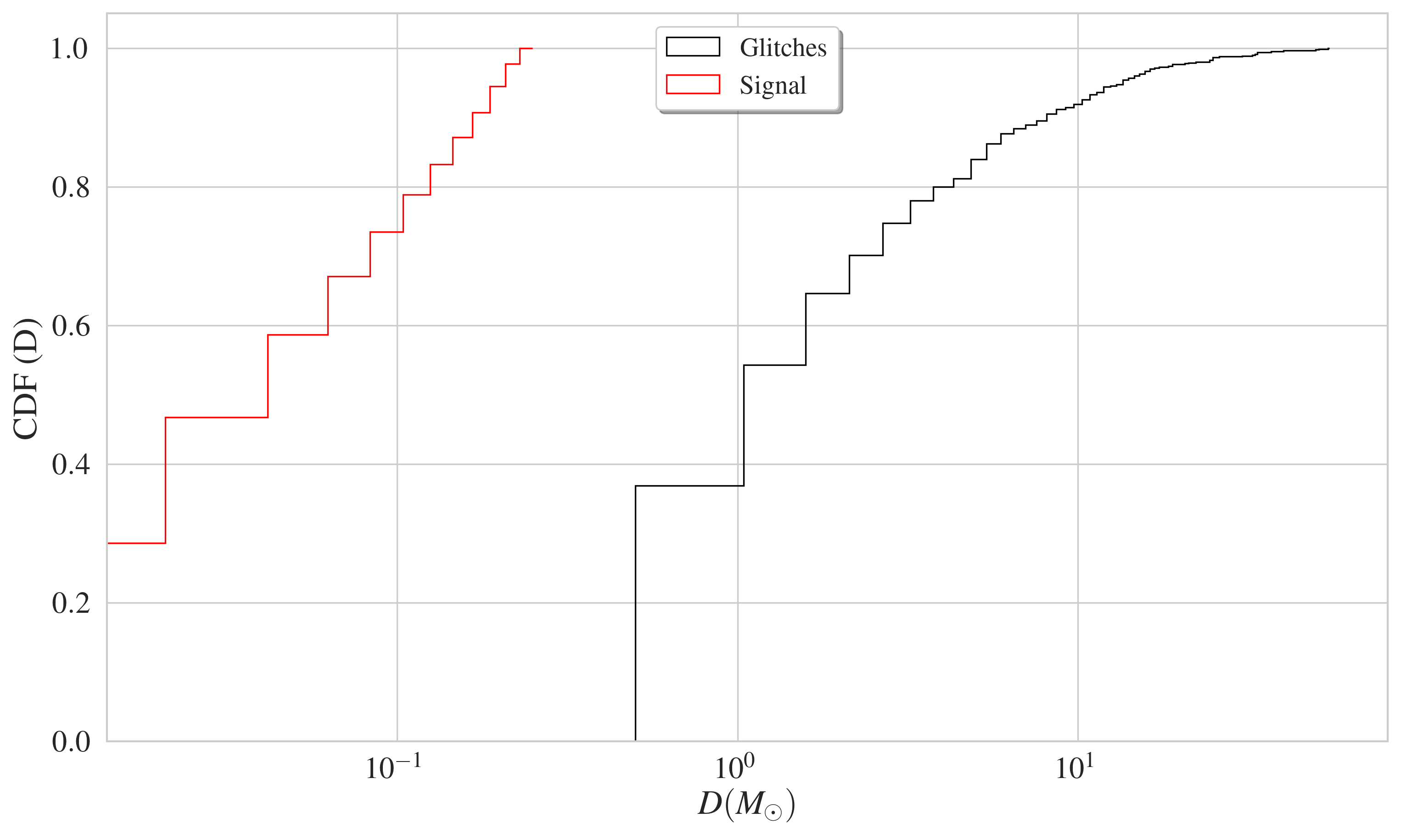}
    \caption{Cumulative distribution for D for all the signal and glitch bins. The black(red) distribution represents the glitch(signal) bins.}
    \label{D distribution}
\end{figure}

We consider various types of short duration glitches, 
observed in the O1 and O2 run \cite{Zevin:2016qwy}.  Among the chosen short duration glitches, we have included chirpy blips, fat blips, tiny blips, raindrop glitches, blips with horns, tomte glitches, and triangular blips, with SNR in the range 9.7 to 47.8. We simulate 1000 compact binary sources with ${\cal M} \le 5 M_{\odot}$ (which translates to signal with a minimum duration of 3 seconds for a low frequency cut off of the detector to be 15 Hz) 
and component masses from 1.4 $M_{\odot}$ with SNR ranging between 10 to 50 with the approximant IMRPhenomPv2~\cite{Husa:2015iqa} and then inject in the data at random location with respect to a glitch.  

We apply an energy cut of 5$\sigma$ and a frequency cut of $\pm$5$\sigma$. We estimate the chirp mass for three Q planes viz. $Q_{best}$ and  $ Q_\pm$ for each time bin. We plot the distribution of D for both glitches and signals in Fig.~\ref{D distribution}.

The black (red) distribution denotes the D distribution for the glitch(signal) bins, respectively. For the signal, the distribution is  confined to the value of $D=0.2 M_\odot$ whereas for the blip $D$ values start from $D=0.5 M_\odot$. This wide spread in the D distribution is due to a wide variety of the glitches considered for the study. Owing to the distinct and separate distributions,  we choose D to be 0.5 $M_\odot$ to separate the signal from the glitch bins. 

\subsection{Glitch bin location}
\label{Results}
\begin{figure}[h!]
    \includegraphics[width=\columnwidth]{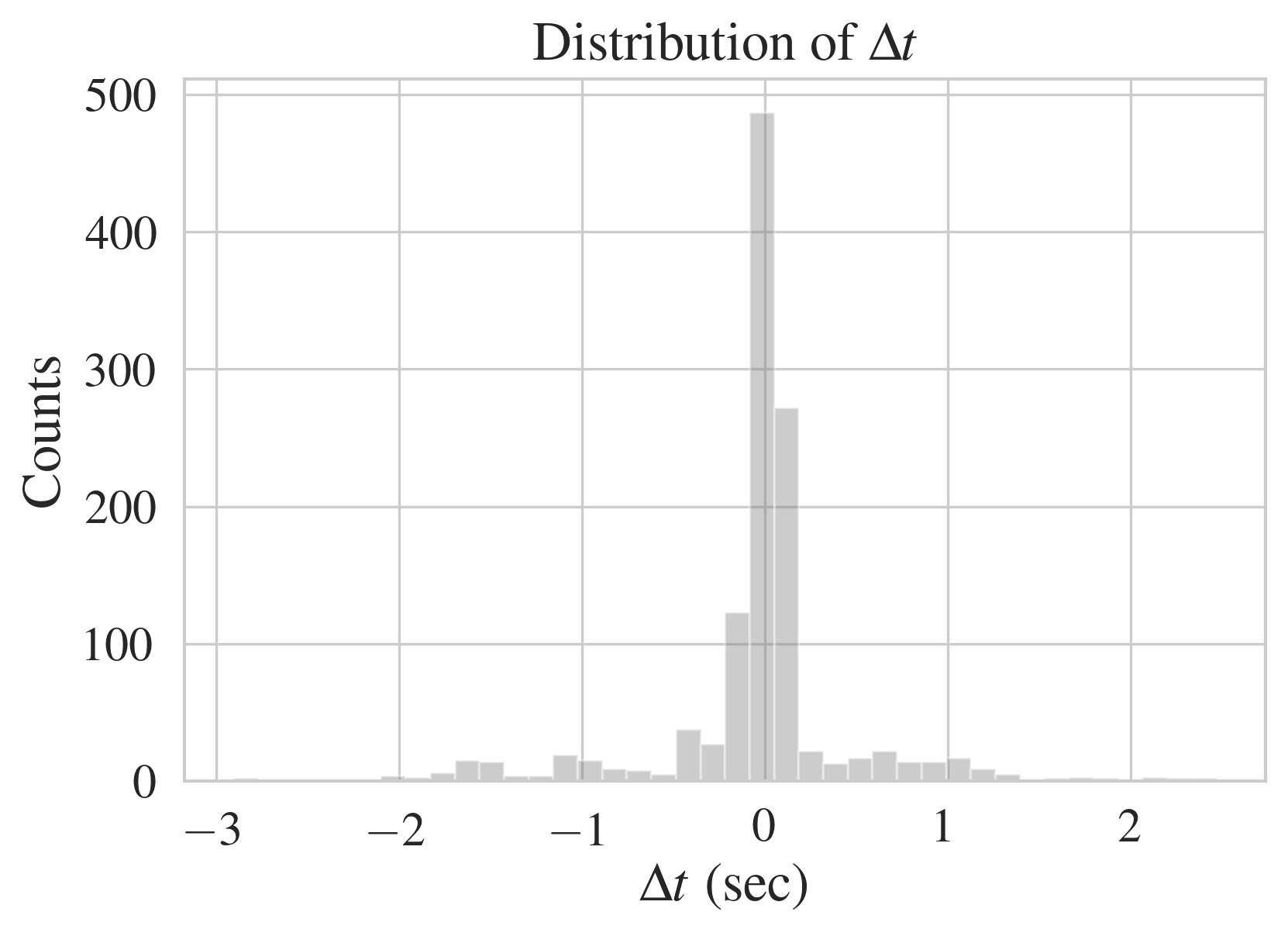}
    \caption{$\Delta t$ distribution for glitches from the simulation study.}
    \label{delta_t counts}
\end{figure}

In addition of the value of D which characterises the morphology, finally we apply the maximum pixel criterion to identify glitch bin.
The chirp mass consistency, along with the maximum pixel bin, addresses cases with low SNR signals and helps to distinguish the merger phase from the noisy glitch. We plot the distribution of $\Delta t = t_i - t_r$ for all the injections to identify the glitch location. Here $t_i$ is the actual location of the glitch, and $t_r$ is the recovered position of the glitch. This distribution, as shown in Fig.\ref{delta_t counts} peaks at $\Delta t = 0$ indicating that for most of the cases ($\sim 90 \%$), the algorithm can recover the glitch bin within an accuracy of less than 0.3 seconds. 
The tail cases where the algorithm deviates from the glitch bin correspond to low blip SNR ($\leq 10.5$)  cases, as shown in Fig \ref{blip SNR vs delta t}\blue{a}, which result in a low number of TF bins. For low SNR glitches, a small part of the glitch is picked up while the majority of the blip is missed, which increases $\Delta t$ for that case. 

\section{Application} \label{Application}
We demonstrate this algorithm for two long duration signal cases; {\it first} the BNS merger event GW170817 data containing a loud short duration glitch and {\it second}, low mass, eccentric binaries with low  eccentricities.
 \\
\subsection{GW170817 with glitch}\label{GW170817 with blip case}
\begin{widetext}
\begin{figure*}[hbt!]
\begin{tabular}{cc}
\centering
  \includegraphics[width=88mm,height=6.2cm]{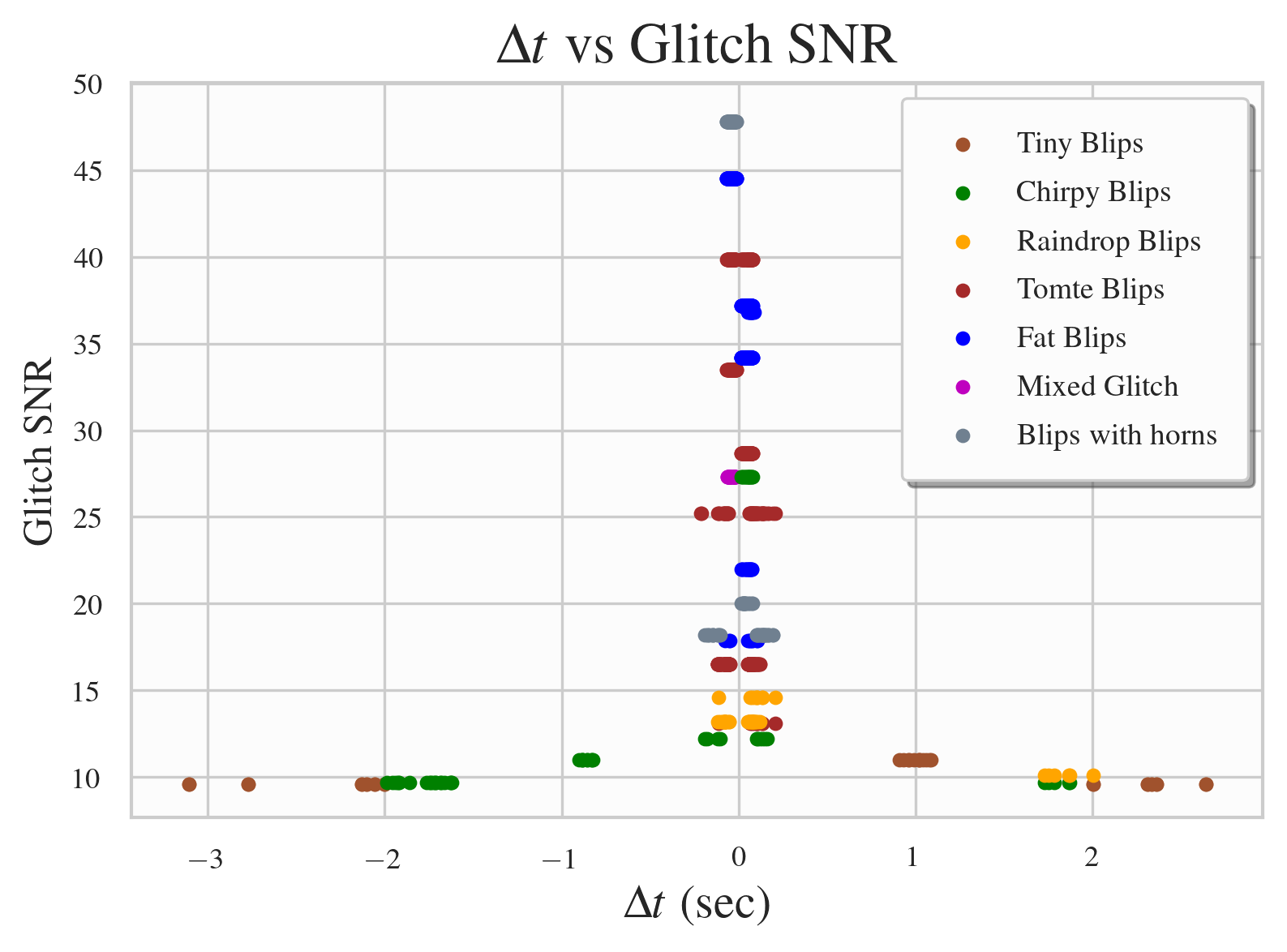} & \includegraphics[width=100mm,height=6.cm]{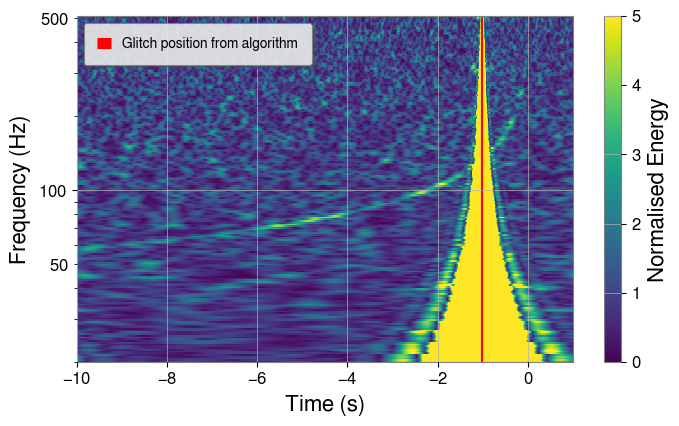}\\ 
  (a) & (b) 
\label{Delta-t}
\end{tabular}
\caption{(a) Variation of $\Delta t$ with glitch SNR. (b) Identification of the glitch from the GW170817 data using our algorithm.}
\label{blip SNR vs delta t}
\end{figure*}
\end{widetext}

The first binary neutron star merger event GW170817 was detected by Advanced LIGO and Virgo detectors with a combined SNR of 32.4. The eccentricity of this event was very low ($\leq 0.024$) \cite{Lenon:2020oza}. A short duration loud glitch was recorded $\sim1.1s$  before the end time of the signal in the LIGO Livingston detector. As a result, the event was identified as a single detector event. This glitch was short duration ($\leq 5 ms$), with a large amplitude spanning over a wide range of frequency, as shown in Fig.~\ref{GW170817 with blip case}. Fig.~\ref{blip SNR vs delta t}\blue{b} shows the LIGO-Livingston data, along with the  the location of glitch to an accuracy of 0.01 second from the proposed algorithm. The red line shows the position of the glitch, as identified by this algorithm.  
 \subsection{Eccentric binaries with low eccentricity}\label{Eccentric binaries with low eccentricity}

In this subsection, we apply this algorithm to identify glitch position in the time-frequency track of a mildly eccentric, low mass binary systems.

We simulate 300 eccentric signals having chirp mass less than 5$M_{\odot}$ using EccentricFD waveform~\cite{PhysRevD.90.084016}. The optimal SNR of them are between 10 to 50 and the eccentricity is chosen randomly from a uniform distribution between 0.1 and 0.4. The waveform constraints restrict the upper limit of eccentricity. These simulated signals are then injected in the LIGO open data of second observing run in a time window containing a real glitch, as discussed in section~\ref{Validation}.  We apply our algorithm to identify the location of the glitch if present in the data. For majority($\sim 87 \%$) of the cases the glitch position is obtained with an accuracy of less than $0.3 $  seconds. 
For relatively higher eccentricity ($>$0.3) and high SNR ($>$40) systems, the recovered glitch position is misestimated because for higher eccentricities and high signal SNR, the higher order eccentric modes start to have non negligible contribution compared to the quadrupolar mode. For such systems, the algorithm requires suitable modification, which includes higher-order modes in the GW signal. We plan to study this in the future.

\section{Conclusions}
\label{Conclusions}
We have constructed an algorithm to identify the position of the short duration glitch if present in the long-duration gravitational wave transients. This method is motivated by the difference in morphology of glitch and long-duration GW signal (including eccentric low mass binaries) in a Q-transform. 
In particularly, the algorithm uses the chirp mass inconsistency across different Q planes as a criterion to identify the glitch location if underlying GW signal. We used energy as well as frequency cut to eliminate the noisy pixels and select the excess energy pixels for estimation of chirp mass parameter. Further, we use the maximum pixel criterion to localize the glitch. 

The algorithm was validated by injecting several simulated low mass GW signals superimposed with short duration glitches mostly blips. The algorithm was also applied on the low mass eccentric binary systems. For most cases, the glitch was accurately identified, including the GW170817 event case. The performance of the algorithm suffered for systems with low glitch SNR. This can be eventually incorporated in the unmodeled eBBH searches which are based on excess power methods.

This method can be suitably extended to any long-duration transient search. Once the glitch is identified in the long duration search, its contribution can then be mitigated by the glitch mitigation methods \cite{Pankow:2018qpo} which will reduce the noise background.\\

\acknowledgements
The authors would like to thank Tito Dal Canton for providing useful comments on the draft. The authors are also grateful to LIGO DetChar group, especially Jess Mclver, Laura Nuttall and Erik Katsavounidis for helpful discussions. The authors acknowledge Gravitational Wave Open Science Center (https://www.gw-openscience.org), a service of LIGO Laboratory, the LIGO Scientific Collaboration and the Virgo Collaboration for usage of open data and softwares. NB acknowledges Inspire division, DST, Government of India for the fellowship support. AP acknowledges the SERB Matrics grant MTR/2019/001096 for support. KC acknowledges MHRD, Government of India, for fellowship support. VG acknowledges Inspire division, DST, Government of India for the fellowship support. This document has LIGO DCC No P2000205.

\bibliographystyle{apsrev4-1}
\bibliography{references}

\appendix
\begin{widetext} 
\section{Q-transform of GW Chirp} \label{A1}

This appendix computes the Q-transform of the GW chirp given by Eqn.\eqref{eq:3}. The bi-square window can be reasonably approximated as a Gaussian window function centred at $f=0$ and is defined as:
\begin{equation} \label{Gaussian Window}
    \mathfrak{\tilde{w}}(f,f_0,Q) \equiv \mathfrak{w}_0 \sqrt{\frac{Q}{f_0}} \exp (-(f^2/\delta^2)) \,.
\end{equation}
where $\delta = 2 f_0/Q$ and $\mathfrak{w}_0$ is a constant. To obtain the Q-transform of a Newtonian chirp, we plug in Eqn.~\eqref{Gaussian Window} and Eqn.~\eqref{eq:3} into Eqn.~\eqref{eq:1}.
We make a suitable change of variable to $z=f+f_0$. We absorb the Fourier kernel into the effective phase and define:
\begin{equation}
    \Psi(z) \equiv 2\pi z(t_c-t) - 2 \phi_c - \frac{\pi}{4} + \frac{3}{128} (\pi G \mathcal{M} z/c^3)^{-5/3} \,.
\end{equation}
The term $\frac{z^{-7/6}}{S_n(z)}$ is slowly varying as compared to the exponential around $f_0$ and hence we treat as a constant and evaluate it at $f_0$. We Taylor expand $\Psi(z)$ around $f_0$ as 
\begin{equation}
    \Psi(z) \sim \Psi(f_0) + K (z-f_0) + C (z-f_0)^2
\end{equation}
with $K = \frac{\partial \Psi}{\partial z}\vert_{z=z_0} =  2\pi \left( t_c - t- \tau\left(\frac{f_L}{f_0}\right)^{8/3} \right)$
where $t_c$ is the time at coalescence, $\tau$ is the chirp duration in the detector band with the low cut-off frequency of $f_L$ and $\tau \left(\frac{f_L}{f_0}\right)^{8/3}$ is the chirp duration when the signal reaches $f_0$ till the $t_c$.
\begin{equation}
    C = \frac{1}{2}\frac{\partial^2 \Psi}{\partial z^2}\vert_{z=f_0} = \frac{5}{96}f_0^{-11/3}(\pi G\mathcal{M}/c^3)^{-5/3} = \frac{8 \pi}{3} \left(\frac{f_L}{f_0} \right)^{8/3} \frac{\tau}{f_0}\,.
\end{equation}

Based on all the above arguments, we are now left with 
\begin{equation}
\label{absX}
        ||X(t,f_0,Q)|| =  h_0 \mathcal{M}^{5/6} \mathfrak{w}_0 \sqrt{\frac{Q}{f_0}} \frac{f_0^{-7/6}}{S_n(f_0)} I(f_0, Q; {\mathcal M})\,,~~~~ I(f_0,Q; {\mathcal M}) = ||\int_{-\infty}^{\infty} dz ~ e^{-iK(z-f_0) - iC(z-f_0)^2}e^{-((z-f_0)^2/\delta^2)}||
\end{equation}
We make change of variable to $x = \sqrt{1/\delta^2 + iC}~(z-f_0)$ and convert the exponent in to a perfect square. The integral is thus converted in to a Gaussian integral which provides the solution as:
\begin{equation}
I(f_0,Q; {\mathcal M}) = \Bigg\| \frac{\sqrt{\pi}}{\sqrt{1/\delta^2 - iC}} \exp\left({\frac{-K^2}{4(1/\delta^2 + iC)}}\right) \Bigg\| =  \frac{2 \sqrt{\pi} f_0}{(Q^4 + 16 C^2 f_0^4)^{1/4}} \exp  \left({\frac{- f_0^2 K^2 Q^2}{Q^4 + 16 f_0^4 C^2}}\right)  
\end{equation}
Substituting in Eqn.~\eqref{absX}, and using $C f_0^2 = \frac{8 \pi}{3} \left(\frac{f_L}{f_0} \right)^{8/3} \tau f_0$, we get

\begin{equation}
    \|X\|^2 = \frac{h_0^2 \mathcal{M}_c^{5/3} \mathfrak{w}_0^2}{S_n(f_0)^2} \frac{4 \pi Q f_0^{-4/3}}{\Bigg(Q^4 + \frac{16*64 \pi^2}{9} \left(\frac{f_L}{f_0} \right)^{16/3} \tau^2 f_0^2 \Bigg)^{1/2}}  \exp \left({\frac{- K^2 f_0^2 Q^2} {Q^4 + 16 \frac{64 \pi^2}{9} \left(\frac{f_L}{f_0} \right)^{16/3} \tau^2 f_0^2}}\right)
\end{equation}
For high values of Q ($>60$), $Q^4$ dominates over the other term reducing the integral to
\begin{equation}
    \|X\|^2 = \frac{4 \pi h_0^2 \mathcal{M}_c^{5/3} \mathfrak{w}_0^2}{Q} \frac{f_0^{-4/3}}{S_n(f_0)^2}  \exp \left(\frac{- K^2 f_0^2} {Q^2} \right)
\end{equation}

\section{Q-transform of a sine-Gaussian transient}
\label{appendixB}
We compute the Q-transform of the sine-Gaussian. The Fourier transform of the sine-Gaussian is given as~\cite{Canton_2013}:
\begin{equation}
    \tilde{g}(f)=\frac{g_0 \sqrt{\pi}}{2}\tau_{sg} \Big([e^{-\pi^2 \tau_{sg}^2(f-f_{sg})^2} \exp[-2 \pi i (f-f_{sg}) t_{sg}] + [e^{-\pi^2 \tau_{sg}^2(f+f_{sg})^2} \exp[-2 \pi i (f+f_{sg}) t_{sg}] \Big)  {\rm for}~~\phi_{sg} = 0.
\label{eq:gf}
\end{equation}
where the quality-factor $Q_{sg}= 2 \pi f_{sg}\tau_{sg}$.  We use the Eqn.~\eqref{Gaussian Window} along with $\tilde{g}(f)$ and substitute in Eqn.~\eqref{eq:1}. Please note that in Eqn. \eqref{eq:gf}, the first(second) term contributes for the +(-) frequencies respectively. After whitening the data, the domain of integration in the Q-transform is positive frequencies. Owing to the narrow width of Q-transform window $\mathfrak{\tilde{w}}(f-f_0,f_0,Q)$ from where the contribution to the integration comes, the noise PSD takes the constant value of $S_n(f_0)$.
\begin{equation}
\| X \| \propto \frac{\tau_{sg}}{S_n(f_0)} \sqrt{\frac{Q}{f_0}} \int_{-f_0}^{\infty}  \exp[-2 \pi i f t'] \exp \left[- \pi^2 (f - \Delta)^2 \tau_{sg}^2 - \frac{f^2}{\delta^2} \right] df \,.
\label{eq:Qblip}
\end{equation}
where $t' = t - t_{sg}$ and $\Delta = f_{sg} - f_0$. Completing the square and owing to the narrow Gaussian profile, Eqn.~\eqref{eq:Qblip} is converted in to a Gaussian integral which gives
\begin{equation}
\| X \| \propto \frac{\tau_{sg}}{S_n(f_0)} \sqrt{\frac{Q}{f_0}} \frac{\delta}{\sqrt{(1 + \pi^2 \delta^2 \tau_{sg}^2)}} \exp \left[- \pi^2 \tau_{sg}^2 \Delta^2 \left(1 - \frac{\pi^2 \tau_{sg}^2 \delta^2}{1+ \pi^2 \delta^2 \tau_{sg}^2} \right) \right] \exp\left(-\frac{\pi^2 t'^2 \delta^2}{1 + \pi^2 \delta^2 \tau_{sg}^2}\right)\,.
\label{eq:Qblip2}
\end{equation}
For short duration glitches $\tau_{sg}$ is few ms duration. As $1 >> \delta^2 \tau_{sg}^2$, the Eqn. simplifies to
 \begin{equation}
\| X \| \propto \frac{\tau_{sg}}{S_n(f_0)} \sqrt{\frac{f_0}{Q}}  \exp \left[- \pi^2 \tau_{sg}^2 (f_0-f_{sg})^2 -\pi^2 (t - t_{sg})^2 \delta^2 \right]\,.
\label{eq:Qblip3}
\end{equation}
The expression clearly shows that for a fixed $Q$, the profile peaks when $(f_0,t)$ coincides with $(f_{sg},t_{sg})$ as expected. The value drops with increase in Q. 
\end{widetext}
\end{document}